\documentstyle[12pt]{article}
\begin{document}
\title{LINEARLY POSITIVE HISTORIES:
PROBABILITIES FOR A ROBUST FAMILY
OF SEQUENCES OF QUANTUM EVENTS
\thanks{Alberta-Thy-43-93, gr-qc/9403055}}
\author{Sheldon Goldstein\\
Department of Mathematics, Rutgers University\\ New Brunswick, NJ
08903,
USA\\ Internet: oldstein@math.rutgers.edu
\and
Don N. Page\\
CIAR Cosmology Program, Theoretical Physics Institute\\
Department of Physics, University of Alberta\\
Edmonton, Alberta, Canada T6G 2J1\\
Internet:  don@phys.ualberta.ca}
\date{(1993 Sept. 2, revised 1995 Mar. 3)}
\maketitle
\large
\begin{abstract}
\baselineskip 25pt
	Nonnegative probabilities that obey the sum rules
may be assigned to a much wider family of sets of histories
than decohering histories.  The resulting {\it linearly
positive histories} avoid the highly restrictive
decoherence conditions and yet give the same probabilities
when those conditions apply.  Thus linearly positive histories
are a broad extension of decohering histories.  Moreover,
the resulting theory is manifestly time-reversal invariant.
\end{abstract}
\normalsize
\pagebreak
\baselineskip 14.7 pt

	Recently there has been considerable interest in
finding a formulation of quantum mechanics which yields
for a closed system not only the probabilities of single events,
but also the probabilities of sequences of events, or histories
[1-8].

	Single events may be described by projection
operators $P$, which are hermitian idempotent
operators, $P=P^{\dagger}=P^2$.
When the quantum state of the closed system is given by the
pure (for the moment) normalized positive-semidefinite
hermitian density matrix $\rho=|\psi\rangle\langle\psi|=\rho^2$,
the probability of the event is
	\begin{equation}
	p=\:\parallel\!P|\psi\rangle\!\parallel^2\:
	=\langle\psi|P^{\dagger}P|\psi\rangle
	=Tr(P\rho P^{\dagger}), \label{eq:1}
	\end{equation}
or, using the hermitian idempotent property of $P$,
the hermiticity of $\rho$, and the cyclic property of the trace,
	\begin{equation}
	p=\langle\psi|P|\psi\rangle
	=Re\,\langle\psi|P|\psi\rangle
	=Tr(P\rho)
	=Tr(\rho P^{\dagger})
	=Re\,Tr(P\rho). \label{eq:2}
	\end{equation}
Any of these expressions written in terms of $\rho$ rather than
$|\psi\rangle$ may also be applied to a system in a mixed
quantum state $\rho\neq\rho^2$ (but still hermitian,
positive-semidefinite, and normalized,
$Tr\rho=1$).

	Although it is doubtful that they can be empirically tested
in general \cite{9}, it is often desirable to be able to assign
probabilities to sequences of events, which can be described by
the class operator
	\begin{equation}
	C=P^{(n)}P^{(n-1)}\cdots
	P^{(2)}P^{(1)}, \label{eq:3}
	\end{equation}
or to more general histories, described by $C$'s
that are sums of such strings (\ref{eq:3}) of projection operators.
Here the projection operators $P^{(i)}$ are projection operators at
different times $t_i$, $t_1<t_2<\cdots<t_{n-1}<t_n$,
and we are using the Heisenberg picture.
When all of the projection operators commute, $C$
itself is a projection operator, and Eqs. (\ref{eq:1}) and
(\ref{eq:2}) would apply with $P$ replaced by $C$.
However, generically when the projection operators do not
commute, $C$ is not a projection operator.  Then the
different expressions in Eqs. (\ref{eq:1}) and (\ref{eq:2})
with $P$ replaced by $C$ may
differ, and it becomes arbitrary which, if any, to use.

	The usual choice [1-8] is to take
the analogue of Eq. (\ref{eq:1}) and say
	\begin{equation}
	p=Tr(C\rho C^{\dagger}). \label{eq:4}
	\end{equation}
This choice has the positive feature that it is always nonnegative.
It is motivated by the fact that for single strings (\ref{eq:3}),
this expression indeed gives the probability of observing the
events $P^{(i)}$ in a sequence of ideal measurements on
a system with initial density matrix $\rho$.
This is a routine consequence of applying the standard
quantum formalism, including the collapse postulate,
to the sequence of measurements. It also follows from
regarding the measured system as a subsystem of a larger
system, one in which, at successive times $t_i$, the
subsystem is coupled to recording devices in such a way
that a sequence of ideal quantum nondemolition measurements
is performed, with the results stored in commuting records.
Then either Eq. (\ref{eq:1}) or (\ref{eq:2}) applied to
the total system, with the projection operator there
being the product of the commuting projection
operators for the records, gives the same answer as
Eq. (\ref{eq:4}) applied to the subsystem treated as isolated.

	However, we are now interested instead in assigning
probabilities to a history of a closed system, such as the
universe, which is not being measured by any external device.
In this case Eq. (\ref{eq:4}) cannot be derived from Eq. (\ref{eq:1})
or (\ref{eq:2}) for a larger system.  Instead, it must simply be
postulated as a new formula.

	A problem with Eq. (\ref{eq:4}) is that it generally does
not give a set of probabilities obeying the standard sum rules.
For an exhaustive set of histories $\{\alpha\}$, meaning that
the sum of the corresponding class operators $C_{\alpha}$
is the identity,
	\begin{equation}
	\sum_{\alpha}C_{\alpha}=I,  \label{eq:5}
	\end{equation}
one can form a coarser-grained set of histories $\{\widehat\alpha\}$
by grouping together the $\alpha$'s into a set of a smaller
number of exclusive and exhaustive $\widehat\alpha$'s.
The class operators for this coarser-grained set are obtained
by summing,
	\begin{equation}
	C_{\widehat\alpha}=\sum_{\alpha\in\widehat\alpha}C_{\alpha},
	\label{eq:6}
	\end{equation}
and then Eq. (\ref{eq:4}) leads to a new set of probabilities
	\begin{equation}
	p_{\widehat\alpha}=Tr(C_{\widehat\alpha}\rho
	C_{\widehat\alpha}^{\dagger}).
	\label{eq:7}
	\end{equation}
But now the trouble is that generically we do not have the
probability sum rule
	\begin{equation}
	p_{\widehat\alpha}=\sum_{\alpha\in\widehat\alpha}p_\alpha.
	\label{eq:8}
	\end{equation}

	A necessary and sufficient condition [1-3] that the sum rule
(\ref{eq:8}) does hold for probabilities defined by Eqs. (\ref{eq:4})
and (\ref{eq:7}) is that
	\begin{equation}
	Re\,Tr(C_{\alpha}\rho C_{\alpha'}^\dagger)=0
	\mbox{ for all pairs }\alpha\neq\alpha',  \label{eq:9}
	\end{equation}
which is called the weak decoherence condition \cite{3}.  It is also
closely related, but not identical, to the consistency condition
\cite{1,2} that was proposed earlier.  In the consistent histories or
decoherent histories approach to quantum mechanics [1-8], one only
assigns probabilities, by Eq.  (\ref{eq:4}), to consistent or
decohering sets of histories $\{\alpha\}$ such that Eq. (\ref{eq:9})
or a slightly different version of it is exactly or approximately
true.  These probabilities then obey the usual rules, but they are
only defined for highly restrictive sets of histories.

	Here we propose a different new formula for the probabilities
of histories in the quantum mechanics of a closed system, namely the
analogue of the last expression of Eq. (\ref{eq:2}):
	\begin{equation}
	p_\alpha
	=Re\,\langle\psi|C_{\alpha}|\psi\rangle
	=Re\,Tr(C_{\alpha}\rho). \label{eq:10}
	\end{equation}
Because this is linear in $C_{\alpha}$, it obviously obeys the
probability sum rule (\ref{eq:8}) when
	\begin{equation}
	p_{\widehat\alpha}=Re\,Tr(C_{\widehat\alpha}\rho).
	\label{eq:11}
	\end{equation}
Therefore, these may be called {\it linear probabilities}.

	The obvious problem with Eq. (\ref{eq:10}) is that it can be
negative.  Therefore, we impose the {\it linear positivity condition}
	\begin{equation}
	Re\,Tr(C_{\alpha}\rho)\geq 0
	\label{eq:12}
	\end{equation}
for all $\alpha\in\{\alpha\}$. Such a set of histories $\{\alpha\}$
obeying the inequality (\ref{eq:12}) will be called a {\it linearly
positive set of histories}.  A member of such a set may then be
called a {\it linearly positive history} (or {\it positive history}
for short).

	Because the linear positivity condition (\ref{eq:12})
for a given state $\rho$ depends only on the $C_{\alpha}$
for each history in question, one can say whether or not a history
is positive without also specifying in which set of histories
it belongs.  This is one immediate way in which
the linear positivity condition is simpler than the weak
decoherence condition (\ref{eq:9}), since the latter depends
not only on the $C_{\alpha}$ of the history in question,
but also on the $C_{\alpha'}$ of all other histories in the set.
This dependence on the complete set of histories in the decohering
case leads to the complication there of needing to consider the
entire set before one can say whether any individual
history is decohering, a complication that is entirely avoided
for positive histories.  (One could define an individual weakly
decoherent history as one for which the minimal set, given by
$C_{\alpha}$ and $C_{\alpha'}=I-C_{\alpha}$, is weakly
decoherent.  Every history in a weakly decoherent set
of histories is an individual weakly decoherent history, but
a complete set of more than two such individual weakly
decoherent histories is generically not weakly decoherent,
whereas any complete set of individually linearly positive
histories is automatically positive.)

	One can readily see that if the system is in a pure state,
and if the class operator $C_{\alpha}$ is a product of rank-one
projection operators onto a succession of pure states,
$Tr(C_{\alpha}\rho)$ is a product of transition amplitudes that start
and end at the system state.  If this product is nonzero, its phase
is Berry's phase \cite{10} for the closed circuit in the projective
Hilbert space that follows the geodesic segments joining the
successive states.  Thus in this special case the linear positivity
condition (\ref{eq:12}) is the condition that the corresponding
Berry's phase is in the first or fourth quadrant (or at its edge).

	Decoherent histories can be given a
time-symmetric generalization motivated by \cite{ABL}
with both initial and final density matrices $\rho_i$ and $\rho_f$
(still hermitian and positive-semidefinite, but no longer
necessarily normalized) \cite{5}.  Then the
weak decoherence condition (\ref{eq:9}) becomes
	\begin{equation}
	Re\,Tr(\rho_f C_{\alpha}\rho_i C_{\alpha'}^\dagger)=0
	\mbox{ for all pairs }\alpha\neq\alpha',  \label{eq:13}
	\end{equation}
and the probabilities (\ref{eq:4}) become
	\begin{equation}
	p_\alpha=Tr(\rho_f C_{\alpha}\rho_i C_{\alpha}^{\dagger})
	/Tr(\rho_f \rho_i). \label{eq:14}
	\end{equation}

	Similarly, the linear positivity condition (\ref{eq:12})
can be generalized to
	\begin{equation}
	Re\,Tr(\rho_f C_{\alpha}\rho_i)\geq 0,  \label{eq:15}
	\end{equation}
and the resulting linearly positive histories can be assigned
the probabilities
	\begin{equation}
	p_\alpha=Re\,Tr(\rho_f C_{\alpha}\rho_i)
	/Tr(\rho_f \rho_i). \label{eq:16}
	\end{equation}

	In either of these cases, single-state quantum mechanics of
a closed system is the special case
	\begin{equation}
	\rho_f=c_1 I,\;\;\rho_i=c_2 \rho  \label{eq:16b}
	\end{equation}
for any positive real constant numbers $c_1$ and $c_2$.
Even if Eq. (\ref{eq:16b})
does not hold, the linear positivity condition (\ref{eq:15})
and linear probabilities (\ref{eq:16}) for the two-state case
are exactly the same as the analogous condition (\ref{eq:12})
and probabilities (\ref{eq:10}) if we take
	\begin{equation}
	\rho=\rho_i\rho_f/Tr(\rho_i\rho_f),  \label{eq:16c}
	\end{equation}
though this need not give an hermitian density matrix $\rho$
if the hermitian $\rho_i$ and $\rho_f$ do not commute.
Thus the two-state case is in fact a special case of an
even broader generalization of linear positive histories,
applying Eq. (\ref{eq:10}) to an arbitrary operator $\rho$
that need not be hermitian or positive semidefinite,
though it should still be normalized so that the sum of
the linear probabilities is unity.

	Inserting (5) into (10) [or into (16)] and expanding,
one finds [Eq. (14) of \cite{6}] that for weakly decohering
sets of histories, the probabilities assigned by Eqs.
(\ref{eq:4}) and (\ref{eq:10}) [or by (\ref{eq:14}) and
(\ref{eq:16})] are identical.  Since these probabilities
are then all nonnegative, we see that the weak
decoherence condition implies the linear positivity
condition and gives the same probabilities.
Of course, the converse is not true.

	Thus the set of all weakly decohering sets of histories
is a proper subset of the set of all linearly positive sets of
histories.  In fact, the weak decoherence condition
(\ref{eq:9}) or (\ref{eq:13}), being a set of {\it equations}
(for all $\alpha\neq\alpha'$), is true only on a surface in
the set of parameters describing a set of histories.
On the other hand, the linear positivity conditions
(\ref{eq:12}) or (\ref{eq:15})
are merely inequalities and so are true in a region
(the closure of an open region)
of the set of parameters describing a set of histories.
That is, the set of all weakly decohering sets of histories
is a subset of measure zero of the set of all linearly positive
sets of histories, whereas the latter is a subset of positive
measure in the set of all sets of histories.

	In this way linearly positive histories are an enormous
generalization of weakly decohering histories.  The former
enable one to assign sets of probabilities to a much broader
family of sets of histories, avoiding the highly restrictive
conditions (\ref{eq:9}) or (\ref{eq:13}) of weakly
decohering histories.  It is also obviously true that linearly
positive histories are an even greater generalization of
histories that obey the medium decoherence condition
\cite{3}, which is Eq. (\ref{eq:9}) without $Re$ on the
left hand side, or the strong decoherence condition
\cite{3}, which is that there exists a complete set of
orthogonal projection operators $R_{\alpha}$ such that
	\begin{equation}
	C_{\alpha}\rho=R_{\alpha}\rho.  \label{eq:16d}
	\end{equation}

	Because of the strong restrictions imposed by the
equations for the various decoherence conditions, often
these are loosened to approximate equalities \cite{3,4}.
However, this procedure has a certain vagueness or
arbitrariness which is entirely avoided by the precise
inequalities (\ref{eq:12}) or (\ref{eq:15}) of the
linear positivity condition.

	The linear probabilities (\ref{eq:10}) and the linear
positivity condition (\ref{eq:12}) for the case of a single state
$\rho$ have the nice feature that they are automatically
invariant under reversing the order of the projection operators
in $C_{\alpha}$, which replaces it by $C_{\alpha}^{\dagger}$.
The same is true for (\ref{eq:15}) and (\ref{eq:16}) with both
initial and final states $\rho_i$ and $\rho_f$ if they commute.
Similarly, if we define the $CPT$-reversed history
$\widetilde{\alpha}$, represented by the class operator
	\begin{equation}
	\widetilde{C}_{\alpha}=
	\Theta^{-1} C^{\dagger}_{\alpha}\Theta
	\label{eq:17}
	\end{equation}
which takes the $CPT$ conjugates of the projection operators
as well as reversing the order \cite{5,6}, then the linear
probabilities and linear positivity condition are invariant
under this ``time reversal'' in the one-state case if, as usual,
$\rho$ is replaced by its ``time reversal''
$\widetilde{\rho}\equiv\Theta^{-1}\rho\Theta$
or, in the two-state case, if $\rho_i$ and $\rho_f$ are replaced,
respectively, by $\widetilde{\rho_f}$ and $\widetilde{\rho_i}$. In
particular, the linear probabilities and linear positivity condition
are invariant (without any change of state) in the one-state case
if $\rho$ is $CPT$ invariant or in the two-state case if
	\begin{equation}
	\rho_i\rho_f=\widetilde{\rho_f}\widetilde{\rho_i}\equiv
	\Theta^{-1}\rho_f\Theta\Theta^{-1}\rho_i\Theta,
	\label{eq:18}
	\end{equation}
e.g., if $\rho_f=\widetilde{\rho_i}\equiv\Theta^{-1}\rho_i\Theta$,
or, alternatively, if $\rho_i$ and $\rho_f$ commute and are
separately $CPT$-invariant, i.e., if $[\rho_i,\rho_f]=0$,
$\rho_i=\widetilde{\rho_i}$, and
$\rho_f=\widetilde{\rho_f}\equiv\Theta^{-1}\rho_f\Theta$.

	It is perhaps worth emphasizing that a set of histories
defining a sequence of {\it measurements\/} automatically satisfies
not merely the linear positivity condition (\ref{eq:12}) but also the
weak decoherence condition (\ref{eq:9}), when the formulas are
applied to the records of the measurements.  (This is true because
each $C_{\alpha}$ is then a product of projection operators that
commute, namely one projection operator for each independent record
of the corresponding measurement.)  Thus the formulas (\ref{eq:10})
and (\ref{eq:4}) agree in this case.

	In the case of {\it ideal} measurements, we could as well
have considered the measured system projections with which the
records are correlated. Moreover, in this case one gets the same
probability from Eq. (\ref{eq:4}) even if one replaces the
$C_{\alpha}$ that is the product of commuting projection operators
onto the records with the $C_{\alpha}$ that is the product of the
corresponding (generically noncommuting) projection operators onto
the measured system {\it treated as closed\/}, i.e., with the
Heisenberg projections defined in terms of the unitary evolution of
this system in isolation.  (This is why in ideal cases one can
correctly calculate the probabilities by an analysis of the measured
system alone, ignoring the quantum mechanics of the measuring
apparatus.)  However, if this replacement is made for each
$C_{\alpha}$, then Eq. (\ref{eq:10}) does {\it not} generically give
the same answer as Eq. (\ref{eq:4}), even when the  histories are
still linearly positive.  Thus the probabilities of linearly positive
histories depend crucially on what measurements are actually made.

For example, consider a spin-${1 \over 2}$ system with
$\rho=|\sigma_z=1\rangle\langle\sigma_z=1|$,
$P^{(1)}_1=|\sigma_x=1\rangle\langle\sigma_x=1|$,
$P^{(2)}_1=|\sigma_z=1\rangle\langle\sigma_z=1|$, and
$P^{(i)}_2=I-P^{(i)}_1$.  The corresponding set of histories, with
elementary class operators (using time-ordered
labeling) $C_{11}=P^{(2)}_1 P^{(1)}_1,\;
	C_{21}=P^{(2)}_1 P^{(1)}_2,\;
	C_{12}=P^{(2)}_2 P^{(1)}_1,\;
	C_{22}=P^{(2)}_2 P^{(1)}_2$, is not weakly
decoherent---because of the obvious interference---but it {\it is\/}
linearly positive, with probabilities $p_{11}=p_{21}=1/2,\quad
p_{12}=p_{22}=0$.   However, if a  measurement of the
first spin (i.e., of $\sigma_x$ at time $t_1$) is incorporated into
the
histories, the resulting set of histories  {\it will\/} be weakly
decoherent, and the probabilities will all become 1/4.

	We note finally that for a different category of histories
than the category [of all histories of the form (3)] considered in
this paper, namely, the category of histories each of which is given
by a collection of (fine-grained) trajectories in configuration space
alone, it is possible to extend the formula (1) applied to
configurational events at any single time to a probability
distribution on the set of {\it all\/} possible configurational
trajectories \cite{Bohm}.  One may thus wonder whether an even
broader extension [of (4) applied to weakly decohering histories]
than that provided by (10) applied to linearly positive histories is
possible, an extension which consistently assigns probabilities to
{\it all\/} possible histories (3).  We note in this regard that such
an extension is precluded by the usual no-hidden-variables theorems
\cite{Bell}.  [These theorems show, in fact, much more: that it is
even impossible to have an extension, to all histories, of (4)
restricted to histories for which the projections in the sequence (3)
mutually commute.]  In other words, the totality of different weakly
decohering sets of histories, or of different linearly positive sets
of histories, with their respective probability formulas, is
genuinely inconsistent---in the sense that the ``probability''
assignments for these different sets of histories cannot
simultaneously be realized as relative frequencies within a single
ensemble.  This shows, in fact, that whatever may be the virtues of
the linear positivity condition, within the framework considered here
it cannot eliminate the necessity, emphasized by Gell-Mann and Hartle
\cite{3,4}, of formulating additional conditions on sets of histories
which select from this totality a limited number of sets of
histories, and perhaps a unique set of histories (e.g., one which
defines what Gell-Mann and Hartle call the ``quasiclassical domain of
familiar experience'' \cite{3}).

	To summarize, linear probabilities (\ref{eq:10})
or (\ref{eq:16}) may be applied to a much broader class of
histories than weakly decohering histories.  That is, they may
be applied to our proposed linearly positive sets of histories,
which are sets of histories obeying the linear positivity condition
(\ref{eq:12}) or (\ref{eq:15}), namely the condition that the
linear probabilities are all nonnegative.  These linear probabilities
obey the sum rules and are equal to the previously proposed
probabilities (\ref{eq:4}) or (\ref{eq:14}) in the very special
subset of cases obeying the weak
decoherence condition (\ref{eq:9}) or (\ref{eq:13}) necessary
for the probabilities (\ref{eq:4}) or (\ref{eq:14}) also to obey
the sum rules.

	We appreciate discussions with James Hartle, Tom\`{a}\v{s}
Kopf, Pavel Krtou\v{s}, Joel Lebowitz, and William Unruh.  This work
was supported in part by NSF Grant No. DMS-9305930 and by NSERC.

\end{document}